\documentclass[preprint,showpacs,preprintnumbers,amsmath,amssymb]{revtex4}

\usepackage{graphicx}
\usepackage{dcolumn}
\usepackage{bm}

\def\be{\begin{equation}}
\def\ee{\end{equation}}

\begin{document}

\title{Dvali-Gabadadze-Porrati Cosmology  in Bianchi I brane}

\author{Rizwan Ul Haq Ansari}
\email{ph03ph14@uohyd.ernet.in}
\author{P K Suresh}
 \email{pkssp@uohyd.ernet.in}
\affiliation{ School of Physics, University of Hyderabad.
Hyderabad 500046.India }


\begin{abstract}
The dynamics of Dvali-Gabadadze-Porrati Cosmology (DGP) braneworld
with an anisotropic brane is studied. The Friedmann equations and
their  solutions are obtained for two branches of anisotropic DGP
model. The late time behavior in DGP cosmology is examined in the
presence of anisotropy which shows that universe enters a
self-accelerating phase much later compared to the isotropic case.
The acceleration conditions and slow-roll conditions for inflation
are obtained.

\end{abstract}


\maketitle
\section{Introduction}
Cosmological models inspired by higher dimensional theories have
received a lot of attention in recent times. In these models our
observed  four-dimensional (4D) universe is a three dimensional
hypersurface (brane) embedded in a  higher dimensional space-time
called bulk (for a review see \cite{brax}). Two popular
cosmological models emerging out of higher
 dimensional theories are Randall-Sundrum (RS) \cite{lisa} and the Dvali-Gabadadze-Porrati (DGP) model
\cite{dvali}. The generalisation of such models to
 the homogenious and isotropic
Friedmann-Robertson-Walker (FRW) brane leads to modification of
 the Friedmann equation with a quadratic correction to
 energy density at higher energies. Many issues in cosmology like inflation, dark
 energy, cosmological constant were investigated in the braneworld cosmological scenario  and encouraging results
 were obtained.

  Although the present universe appears homogeneous and isotropic  in its overall structure, as indicated from recent WMAP data that
  cosmic microwave background is isotropic up to 1 part in $10^5$. But there are reasons
to believe that it has not been so in all its evolution and that
inhomogeneities and anisotropies played an important role in the
early universe. There exists a large number of  anisotropic
cosmological models,
 which are also being studied in cosmology, due to various reasons \cite{mis}.

Hence, it is natural to ask how these anisotropies play a role in
the context of the braneworld scenario. The anisotropic braneworld
with a scalar field is studied
 in \cite{varun} and it is shown that a large initial anisotropy
 introduces more damping in the scalar field equation of motion resulting in greater inflation. Cosmological solutions for the Bianchi-I and
 Bianchi-V in the case of the RS branemodel were studied in \cite{camp}. It is shown that for
 matter on the brane  obeying the bariotropic equation of state, the anisotropic Bianchi-I,V braneworlds always isotropise although there could be intermediate
 stages in which anisotropy grows. The shear dynamics in the Bianchi-I brane model were studied in \cite{top} and
 shown that for $1<\gamma<2$ shear has maximum value during the phase transition from nonstandard to standard
  cosmology. The cosmological solution of  field equations for an
  anisotropic brane in the generalised RS model were obtained in \cite{pal}
  and the  solution admits an  inflationary era. Dissipation of anisotropy on the brane is explicitly demonstrated by the particle production mechanism  in \cite{gus} by
  considering the Bianchi-I braneworld model. Apart from this there have
  been studies \cite{frolov} which  show that anisotropy plays an important role in braneworld models. All the previous  studies have mainly focussed on anisotropy on the
  Randall-Sundrum braneworld.

   In this paper we consider the DGP braneworld model
  with anisotropy and find a solution of field equations. The DGP cosmological  model
  possesses   two classes of solutions; one which is close to standard FRW cosmology and the  other is a fully five dimensional regime or a
self-inflationary solution which produces  accelerated expansion.
It is also noted in \cite{ant} that the self-accelerating universe
exhibit ghost  and tachyonic like excitations. We  are mainly
interested in
 effects of initial anisotropy and show that it is possible to get self-inflationary solution in the presence of anisotropy. We also
 explore an acceleration condition for the DGP cosmological model with a scalar field dominated
 universe,
in the presence of shear.
\section{Field equations in the DGP model}
We consider the DGP model, where our universe  is a 3-brane
embedded in 5D bulk with an infinite size extra dimension and
there is an induced 4D Ricci scalar on the brane, due to radiative
correction to the graviton propagator on the brane. In this model
there exists a length scale below which the potential has usual
Newtonian form and above which the gravity becomes five
dimensional. The crossover scale between the four-dimensional and
five-dimensional gravity is given by,
 \be
 r_c=\frac{k_5^2}{2\mu^2}
 \ee
 where $\mu^2 = 8 \pi G_4 $ and $k^2_5= 8 \pi G_5$ are the constants related to 4D and 5D Newton's constants respectively.
 The Einstein equation for a five dimensional bulk is given by,
\begin{equation}
G_{AB}= -\Lambda g_{AB}+ k_5^2 (T_{AB} + S_{AB} \delta(y)),
\end{equation}
where $\Lambda$ is the bulk cosmological constant, $T_{AB}$ is the
five-dimensional energy momentum tensor (A,B = 0,..4 ) and
$S_{AB}$ is the 4D energy-momentum tensor and is given by,
\begin{equation}
S_{AB}=\tau_{AB}-\sigma q_{AB} -\mu^{-2} G_{AB},
\end{equation}
where $\tau_{AB}$ is the energy-momentum tensor of matter fields
on the brane, $\sigma$ is the brane tension, and the last term in
the above equation represents contribution from the induced
curvature term. Assuming $Z_2$ symmetry for the brane and using
the Israel junction condition, the effective Einstein equation on
the brane is obtained as \cite{sas,riz},
\begin{eqnarray}\label{gendgp}
 \left( 1+\frac{\sigma k^2_5}{6 \mu^2}\right )
G_{\mu\nu}&=&-\left(\frac{k^2_5
\Lambda}{2}+\frac{k^4_5\sigma^2}{12}\right
)q_{\mu\nu}+\mu^2 T_{\mu\nu} \nonumber  \\
 &&     +\frac{\sigma k^4_5}{6}\tau_{\mu\nu} +\frac{k^4_5}{\mu^4}
F_{\mu\nu} + k^4_5 \Pi_{\mu\nu} + \frac{ k^4_5}{\mu^2}
L_{\mu\nu}-E_{\mu\nu},
\end{eqnarray}
\begin{equation}
\Pi_{\mu\nu} = -\frac{1}{4}\tau_{\mu\rho}\tau^\rho_\nu
+\frac{1}{12}\tau \tau_{\mu\nu}+\frac{1}{8} q_{\mu\nu}
\tau_{\alpha\beta}\tau^{\alpha\beta}-\frac{1}{24}q_{\mu\nu}
\tau^2,
\end{equation}
\begin{equation}
F_{\mu\nu} = -\frac{1}{4}G_{\mu\rho}G^\rho_\nu+\frac{1}{12} G
G_{\mu\nu}+\frac{1}{8} q_{\mu\nu}
G_{\alpha\beta}G^{\alpha\beta}-\frac{1}{24}q_{\mu\nu} G^2,
\end{equation}
\begin{equation}
L_{\mu\nu} = \frac{1}{4}(G_{\mu\rho}\tau^\rho_\nu + \tau_{\mu\rho}
G^\rho_\nu) - \frac{1}{12} (\tau G_{\mu\nu} + G \tau_{\mu\nu}
)-\frac{1}{4} q_{\mu\nu}
(G_{\alpha\beta}\tau^{\alpha\beta}-\frac{1}{3} G \tau),
\end{equation}
 where $T_{\mu\nu}$ is the  bulk energy-momentum tensor. It is
noticed that in Eq. (\ref{gendgp}), apart from the usual quadratic
matter field corrections to energy momentum, there are corrections
coming from the induced curvature term through $F_{\mu\nu}$ and
$L_{\mu\nu}$. $E_{\mu\nu}$ is the projection of the bulk Weyl
tensor on the brane representing the nonlocal effects from the
free gravitational field. If we define the four velocity comoving
with matter as $u_\mu$ \cite{hawk}, the non-local term takes the
form, \be\label{weyl}
 E_{\mu\nu}= -4 r_c^2[U(u_\mu u_\nu +\frac{1}{3}h_{\mu
\nu})+P_{\mu\nu} +Q_\mu u_\nu + Q_\nu u_\mu].
 \ee
where $h_{\mu\nu} = g_{\mu \nu} + u_\mu u_\nu $  and \be \label{U}
U=-\frac{1}{4 r_c^2} E_{\mu\nu}u^\mu u^\nu, \ee is an effective
nonlocal energy density on the brane, \be \label{Pu}
P_{\mu\nu}=-\frac{1}{4 r_c^2} \left[ h_\mu{}^\alpha
h_\nu{}^\beta-{\textstyle{1\over3}}h^{\alpha\beta}
h_{\mu\nu}\right]  E_{\alpha\beta} \ee is the effective nonlocal
anisotropic stress, and
 \be \label{Qu}
  Q_\mu= \frac{1}{4 r_c^2} h_\mu{}^\alpha
E_{\alpha\beta}u^\beta\,, \ee represents the effective nonlocal
energy flux on the brane.

At this stage we set the bulk cosmological constant, brane tension
to zero and consider empty bulk, then  Eq. (\ref{gendgp}) becomes,
\begin{equation}\label{dgpe}
G_{\mu\nu}=\frac{k^4_5}{\mu^4} F_{\mu\nu} + k^4_5  \Pi_{\mu\nu} +
\frac{ k^4_5}{\mu^2} L_{\mu\nu}-E_{\mu\nu}.
\end{equation}
We combine all the nonlocal  and local bulk corrections and these
can be written in an compact form as,
\begin{equation}\label{eins}
G_{\mu\nu}=k^4_5 T^{\rm tot}_{\mu\nu}\,,
\end{equation}
where,
 \be T^{\rm tot}_{\mu\nu}=
\frac{1}{\mu^4}F_{\mu\nu}+\Pi_{\mu\nu}+ \frac{ 1}{\mu^2}
L_{\mu\nu}- {1\over k^4_5} E_{\mu\nu}\,. \ee Following \cite{mar}
the effective  energy density, pressure, anisotropic stress, and
energy density, for the DGP case can be calculated and are
obtained as,
 \be\label{ro} \rho_{\rm tot} =\frac{1}{12}\left(
\rho^2 -\frac{2}{\mu^2}G_{00}\rho +  \frac{1}{\mu^4}G_{00}^2
\right) +\frac{4 r_c^2}{k^4_5} U, \ee
\be\label{p} p_{\rm tot} =\frac{1}{12}\left( \rho(\rho+2p)
-\frac{2}{\mu^2}G_{00}(\rho+p) -\frac{2}{\mu^2}G_{ii}\rho +
\frac{1}{\mu^4} G_{00}(G_{00}+2G_{ii}) \right) +\frac{4 r_c^2}{3
k^4_5} U, \ee
\be\label{pi} \pi^{\rm tot}_{\mu\nu} = \frac{4 r_c^2}{k^4_5}
P_{\mu\nu}, \ee \be\label{q} q^{\rm tot}_\mu = \frac{4
r_c^2}{k^4_5} Q_\mu. \ee It is assumed that the total
energy-momentum tensor and the brane energy-momentum tensor are
conserved independently \cite{mar}.
 It can be seen from Eqs.  (\ref{weyl}-\ref{Qu}) that results depend on the
 crossover scale $r_c$ of the theory, which is a typical feature of the DGP
 model. When $r_c$ goes to zero the effective nonlocal energy density, anisotropic stress, and energy flux diverge.
\section{DGP Cosmology on a Bianchi-I brane}
We consider  an anisotropic Bianchi-I brane model with the metric
given by,
\begin{equation}\label{bianchi1}
ds^2 = -dt^2 + a_i^2(t) (dx^i)^2\,,
\end{equation}
and is covariantly characterized by \cite{varun}
\begin{equation}\label{bianchi}
D_{\mu} f=0\,,~A_\mu=0=\omega_\mu\,,~ Q_\mu=0\,,~R^*_{\mu\nu}=0\,,
\end{equation}
where $D_\mu$ is the projected covariant spatial derivative, $f$
is any physical defined scalar, $A_\mu$ is the four acceleration,
$\omega_\mu$ is the vorticity and $R^*_{\mu\nu}$ is the Ricci
tensor of the three surface orthogonal to $u_\mu$.

The conservation equations  takes the following form:
\begin{equation}\label{cons}
\dot{\rho}+\Theta(\rho + p)= 0
\end{equation}
\begin{equation}
D^\nu P_{\mu\nu} = 0  \end{equation}
 \begin{equation} \label{cons1}
 \dot{U}+{\textstyle{4\over3}}\Theta{ U}+\sigma^{\mu\nu}{
P}_{\mu\nu} = 0
 \end{equation}
where the dot denotes the $u^\nu\nabla_{\nu} $ and $\Theta$
represents the volume expansion rate and $\sigma^{\mu\nu}$ is the
shear rate. The Hubble parameter for the Bianchi-I metric is given
by $H_i = \frac{{\dot a_i}}{a_i}$ and one can define the mean
expansion factor as $S = (a_1a_2a_3)^{1/3}$, thus  \be \Theta
\equiv 3H = 3\frac{{\dot S}}{S} \equiv \sum_{i} H_i.\ee
 The Raychaudhuri equation  for the Bianchi-I metric on the brane is obtained as,
\begin{eqnarray}
\dot{\Theta}+\frac{1}{3}\Theta^2 +\sigma^{\mu\nu}\sigma_{\mu\nu} &
=& \frac{-k^2_5}{2}\left[ \frac{1}{12 \mu^4}\left(3 \left( {
H^2+\frac{K}{S^2}} \right) -\mu^2\rho\right)^2 \nonumber
+\frac{1}{4}\rho(\rho+2p)   \right. \\  && \left.  - \frac{1}{2
\mu^2}\left( 3 \left( { H^2+\frac{K}{S^2}}\right) (\rho+p)
+G_{ii}\rho \right) \right. \nonumber  \\  && \left. +
 \frac{1}{4 \mu^4}3 \left( H^2+\frac{K}{S^2}\right) \left(3\left(  H^2+\frac{K}{S^2}\right)  +2G_{ii}\right) +\frac{8r_c^2}{k^4_5}U \right].
\end{eqnarray}
and Gauss-Codacci equations are \be \label{puv}
\dot{\sigma}_{\mu\nu}+\Theta\sigma_{\mu\nu}=4 r_c^2  P_{\mu\nu}\,
\ee \be\label{cod} \frac{2}{3}\Theta^2
+\sigma^{\mu\nu}\sigma_{\mu\nu}
 = \frac{k^4_5}{6\mu^4}\left(3\left( H^2+\frac{K}{S^2}\right) -\mu^2\rho\right)^2  + 4 r_c^2 U\,.
\ee It is noticed that there is no evolution equation for
$P_{\mu\nu}$, indicating the fact that  in  general the equations
do not close on the brane and complete bulk equations are needed
to determine the brane dynamics. There are bulk degrees of freedom
whose impact cannot be predicted by brane observers. Hence, the
presence of $P_{\mu\nu}$ in Eq. (\ref{puv}) does not allow us to
simply integrate to get the value of shear as in general
relativity. We can overcome this problem by considering a special
case where  nonlocal energy density vanishes or is
negligible\cite{varun} i.e. $U=0$, which is often
 assumed for FRW branes and leads to conformally flat bulk geometry. Under this assumption conservation Eq. (\ref{cons1}) leads to
 $\sigma_{\mu\nu}P^{\mu\nu}=0$;
 this consistency condition implies a condition on evolution of $P_{\mu\nu}$ as  $\sigma^{\mu\nu}\dot{{
P}_{\mu\nu}}= 4 r_c^2 P^{\mu\nu} P_{\mu\nu}$, folowing from Eq.
(\ref{puv}). As there is no evolution equation for $P_{\mu\nu}$ on
the brane \cite{mar,varun}, this is  consistent on the brane.
Also, one should check that the brane metric with $U=0$ leads to a
physical 5D bulk  metric. This has to be done numerically as the
bulk metric for the Bianchi brane is not known  and is beyond the
scope of the present paper.

 Now Eq. (\ref{puv}) can be integrated after contracting with shear \cite{varun} and gives
\begin{equation}\label{s}
\sigma^{\mu\nu}\sigma_{\mu\nu}  = 2\sigma^2 = {6\Sigma^2\over
S^6}\,,~~\dot\Sigma=0\,.
\end{equation}
\begin{center}
\begin{figure*}
\includegraphics{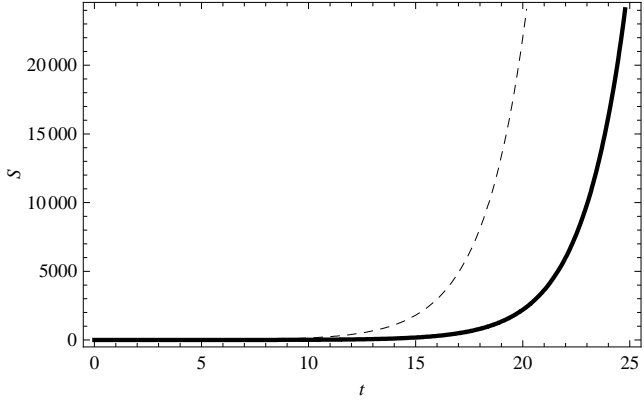}
\caption{Evolution of scalar factor with and without anisotropy,
for $r_c=2 $ and $\Sigma=0.001$ for $\epsilon=+1$. The dashed line
shows the behaviour in the isotropic DGP model and the thick line
corresponds to Bianchi-I DGP brane. It is observed that the
universe enters the self-accelerating phase much later compared to
the isotropic case.}
\end{figure*}
\end{center}
\subsection{Friedmann equation for Bianchi-I  Brane}
The generalised Friedmann  equation in DGP cosmology for the
Bianchi-I brane is obtained using metric (\ref{bianchi1}) and Eqs.
(\ref{cod}, \ref{s}) as
 \be\label{dgp}
 (H^2+\frac{K}{S^2}) -\frac{\Sigma^2}{S^6}-\frac{2 \mu^2}{k^2_5}\epsilon \sqrt{(H^2+\frac{K}{S^2})-\frac{\Sigma^2}{S^6}} =\frac{\mu^2}{3}\rho. \ee
The above equation reduces to the DGP Friedmann equation of
\cite{def} for isotropic universe when $\Sigma=0$. Here,
$\epsilon=\pm 1$ corresponds to two possible embedding of the
brane in the bulk space-time. We can obtain  the Bianchi-I
equation for general relativity  from Eq. (\ref{dgp}) under the
condition \be \label{std}
\sqrt{(H^2+\frac{K}{S^2})-\frac{\Sigma^2}{S^6}}\gg \frac{2
\mu^2}{k^2_5}, \ee which matches the crossover scale set by
\cite{dvali}, but  in this case it also depends on the shear. Also
see that (\ref{std}) reduces to the condition of Deffayet for
recovery of standard cosmology, in the absence of shear.

It can be seen from the Eq. (\ref{dgp}) that we recover the
Friedmann equation of \cite{varun} when $\mu$ goes to infinity. It
also corresponds to the fully five dimensional regime, as $H$ goes
linear to  $\rho$. This can be written equivalently as,
 \be \label{late}
\sqrt{(H^2+\frac{K}{S^2})-\frac{\Sigma^2}{S^6}}\ll \frac{2
\mu^2}{k^2_5}. \ee which is the condition to recover the fully 5D
regime from (\ref{dgp}).
\begin{figure*}
\includegraphics{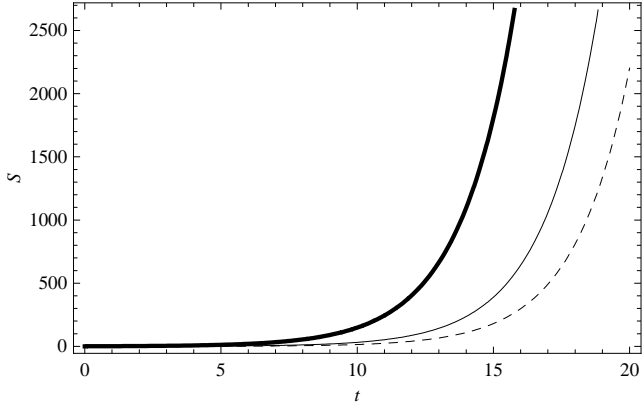}
\caption{Evolution of a scalar factor with different values of
$\Sigma$, thick line ($\Sigma$=1), line ($\Sigma$=0.1) ,  dashed line ($\Sigma$=0.001), for
$\epsilon=+1$ with $r_c=2 $. The
figure shows that the higher value of
 $\Sigma$ implies entering in to the self-inflationary phase much earlier than the smaller value of $\Sigma .$}
\end{figure*}
\subsection{Solution in the late universe}
One of the features of DGP cosmology is that either it enters a
fully 5D regime or a self-accelerating phase, in the late
universe, depending on the value of $\epsilon$. Equation
(\ref{dgp}) can be written as (with $K=0$)
\begin{equation}\label{dgpfred}
   H^2 =\frac{1}{4r_c^2}\left(\epsilon + \sqrt{1 + \frac{4\mu^2 \rho r_c^2}{3}}  \right)
   ^{2} + \frac{\Sigma^2}{S^6}.
\end{equation}
The above equation can be expanded under the condition $ \mu^2
\rho r_c^2 \ll 1$  [following from (\ref{late})],  for the
$\epsilon=1$ case at zeroth order we get,
\begin{equation}\label{late1}
H^2= \frac{1}{r_c^2}+ \frac{\Sigma^2}{S^6}
\end{equation}
which can be compared with the corresponding equation of
\cite{def,gumj} and in the absence of $\Sigma$ matches their
result. Also, for $S\rightarrow\infty$ as expected the shear term
damps and Eq.(\ref{late1}) conforms with \cite{def}. Then the
solution of Eq. (\ref{late1}) is obtained as
\begin{equation}\label{sol1}
S^3= r_c \Sigma \; \hbox{Sinh}( \frac{3t}{r_c}),
\end{equation}
which shows self-acceleration in the late universe; this feature
is already shown  by Friedmann Eq.(\ref{late1}). Hence, it
indicates that the DGP model can  lead to self-inflation even in
the presence of anisotropy. Figure 1 shows behavior of the scale
factor for the anisotropic and the isotropic case. Figure 2 and 4
shows growth of $S$ for different values of $\Sigma$ and $r_c$.

Similarly Eq. ({\ref{dgpfred}) can be expanded  for the $\epsilon
= -1$ case and it gives
\begin{equation}
H^2=  \frac{\Sigma^2}{S^6},
\end{equation}
which shows that the  expansion rate is dominated by shear term.
The solution is,
\begin{equation*}
S^{3}=S_0^{3} + 3 \Sigma (t-t_0).
\end{equation*}
The expansion rate in the present case is slower in comparison to
standard cosmology and the result matches results of \cite{varun}
for the RS type brane model . This is expected in the case of
$\epsilon=-1$ which corresponds to the fully 5D regime and is the
same as the RS model when quadratic term dominates. Figure 3 shows
behavior of $S$ for different values of $\Sigma$.  To discuss the
early universe scenario, we rearrange Eq. (\ref{dgp}) as follows:
\begin{equation}\label{earlydgp}
H^2 =\frac{\mu^2 \rho
}{3}\left(\frac{\epsilon}{2r_c}\sqrt{\frac{3}{\mu^2\rho}}  +
\sqrt{1 + \frac{3}{4r_c^2 \mu^2 \rho }} \right)
   ^{2} + \frac{\Sigma^2}{S^6}.
\end{equation}
At high energies the above equation can be expanded in terms of $
\mu^2 \rho r_c^2 \gg 1$. Thus at zeroth order we get the equation
of the Bianchi-I model of the general relativity.
\subsection{Acceleration conditions}
Next we get the acceleration conditions in  an anisotropic DGP
brane model. Consider Eq. (\ref{dgpfred})  up to first order for
the $\epsilon= +1$ case. We obtain the Friedmann equation,
\begin{equation}\label{late2}
H^2= \frac{1}{r_c^2}+ \frac{2}{3}\mu^2 \rho +
\frac{\Sigma^2}{S^6}.
\end{equation}
The accelerated expansion in standard cosmology  is given by
$\dot{H}+ H^2>0$. Thus using  Eqs. (\ref{cons}) and
eqn(\ref{late2}) the acceleration condition for the present case
becomes
\begin{figure*}
\includegraphics{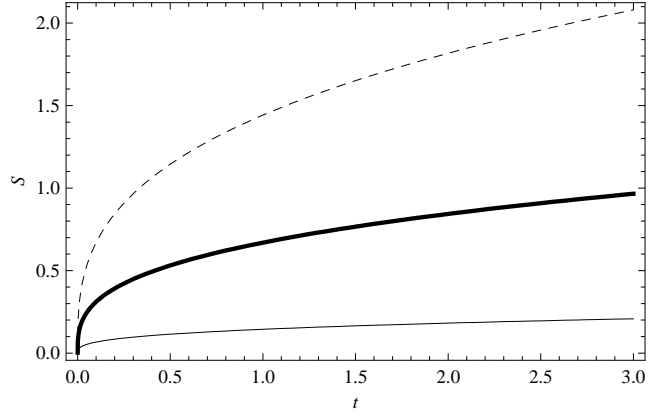}
\caption{Evolution of scalar factor for the case $\epsilon=-1$,
with different values of $\Sigma$. Dashed line ( $\Sigma$=1), thick line
($\Sigma$=0.1), line ($\Sigma$= 0.001).}
\end{figure*}
\begin{center}
\begin{figure*}
\includegraphics{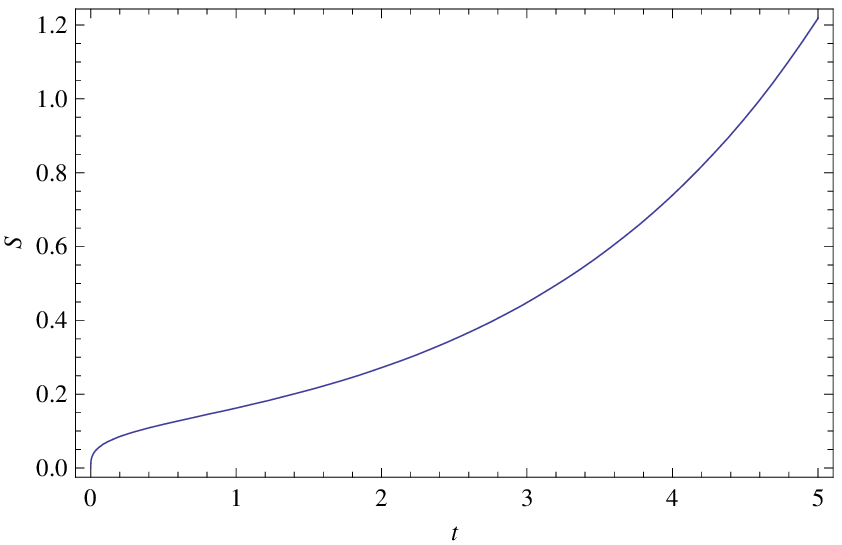}\\
\includegraphics{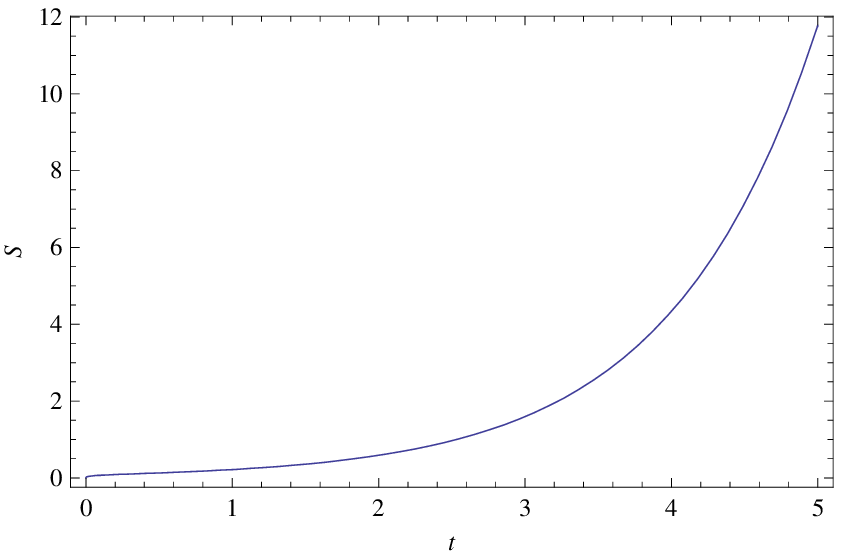}\\
\includegraphics{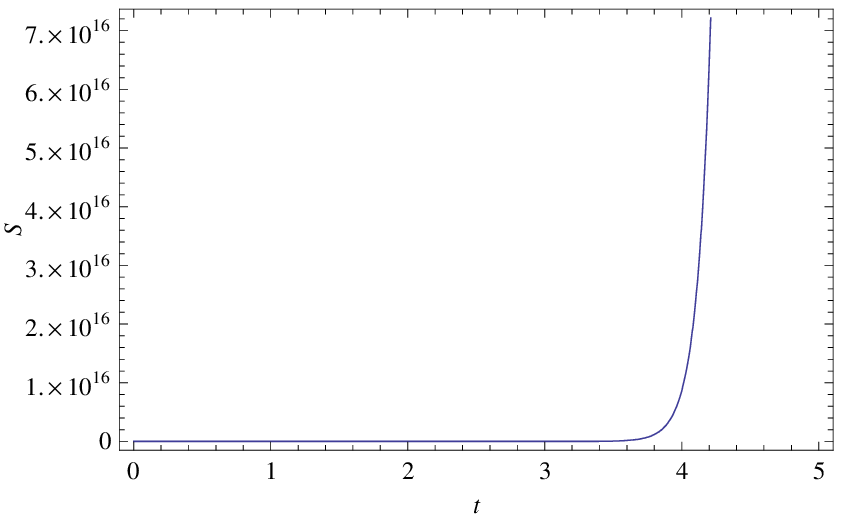}
\caption{Evolution of scalar factor in the $\epsilon=+1$ case,
with different values of $r_c=2,1,0.1 $ for $\Sigma=0.001 $. The
figure shows that scale factor grows faster for small values of
$r_c$,  compaired to higher values.}
\end{figure*}
\end{center}
\begin{equation}\label{accl1}
\dot{H}+ H^2 \simeq -\mu^2(\rho +p) + \frac{1}{r_c^2} +
\frac{2}{3}\mu^2 \rho -2 \frac{\Sigma^2}{S^6} > 0 ,
\end{equation}
and this gives,
\begin{equation}\label{accl3} p < -\frac{\rho}{3} +
\frac{1}{\mu^2 r_c^2 } - \frac{2 \Sigma^2}{\mu^2 S^6}.
\end{equation}
Similarly for the $\epsilon=-1$ case, expanding (\ref{dgpfred})
and considering the next order, the Friedmann equation is,
\begin{equation}\label{fred2}
 H^2=  \frac{\mu^4 r_c^2}{9}
\rho^2 + \frac{\Sigma^2}{S^6}.
\end{equation}
Notice that the relation between the Hubble radius and
 energy density is linear, a feature of brane \cite{sas,lang}, which
is referred to as the fully 5D regime. The acceleration
condition,
\begin{equation}\label{accl2}
\dot{H}+ H^2 \simeq -\frac{\mu^4 r_c^2}{3} \rho(\rho +p)  +
\frac{\mu^4 r_c^2}{9} \rho^2 -2 \frac{\Sigma^2}{S^6} > 0,
\end{equation}
implies
\begin{equation}\label{accl4}
p < -\frac{2\rho}{3}  - \frac{6 \Sigma^2}{S^6\mu^4 r_c^2 \rho }.
\end{equation}
Therefore, one can see that the condition is modified compared to
the standard cosmology and our results reduce to \cite{gumj} in
the absence of anisotropy.

Next we consider the inflationary phase driven by  a scalar field,
with energy density and pressure given by $\rho= \frac{1}{2}
\dot{\phi}^2 + V(\phi)$ and $p = \frac{1}{2} \dot{\phi}^2
-V(\phi)$, respectively. The Klein-Gordon equation for a scalar
field is $ {\ddot \phi} + 3H {\dot \phi} + V'(\phi) = 0$. Using
Eq. (\ref{accl3}) the slow-roll condition for inflation can be
obtained for $\epsilon=1$ case as,
\begin{equation}
 \dot{\phi}^2 < V + \frac{3}{2\mu^2 r_c^2}- \frac{3
 \Sigma^2}{S^6}.
\end{equation}
It can be seen that the  condition depends on the anisotropy and
in  the ordinary DGP conditon is recovered in absence of $\Sigma$
. Similarly for $\epsilon=-1$, using eqn (\ref{accl4})
the slow roll condition is obtained  as,
\begin{equation}
 \dot{\phi}^2 < -4V + 2 \sqrt{3V^2 + \frac{18 \Sigma^2 }{S^6\mu^4 r_c^2}}.
\end{equation}
\section{Conclusions }
 To summarize, we considered anisotropic effects in the DGP  cosmological scenario and
 found a solution to the corresponding field equation. We obtained the Friedmann equations  with the Bianchi-I brane and
  two branches of solutions in the DGP model are considered. The evolution of a scale factor  in the case of the late universe is studied
  and an   acceleration condition for inflation is derived.

 It is observed that, for the $\epsilon=+1$ case, there exists a self-inflationary solution which leads to accelerated
expansion in the late universe, even in the presence of
anisotropy. The evolution of a scale factor is given
 in Fig. 1, which shows that the presence of shear  does not stop the universe from entering a self-accelerated phase, but slows it down. The evolution
 scale factor in the anisotropic DGP model depends on the cross-overscale and the shear. It is noted that, when shear is unity($\Sigma=1$), the evolution of a
 scale factor coincides with the isotropic DGP model.  Fig.2  shows dependence of a scale factor on the shear and the higher the value of $\Sigma$ the
 faster the universe enters into a
  self-accelerated phase. Fig.4 shows the behavior of a scale factor with typical values of  a crossover scale $r_c$ and shows that the higher
  the $r_c$ the slower the acceleration. Hence, in the DGP model shear and $r_c$  behave oppositely  to each other. The behavior $S$ for
  the $\epsilon=-1$ branch of the solution of the Friedmann equation is shown in Fig.3 and a scale factor grows faster for higher values of
  anisotropy. The scale factor grows as
  $S\propto t^{1/3}$ which implies slower expansion than the standard
  cosmology and is similar to expansion rate that of Randall-Sundrum case.
  The Friedmann equation in this case is similar to the RS
  Friedmann equation with quadratic energy density. The acceleration conditions and slow-roll conditions for inflation are obtained and they depend on the shear.

Finally, our results  coincide with those of Deffayet \cite{def}
whenever $\Sigma$ is absent and reduce to the Bianchi-I of the
general relativity under condition  (\ref{std}). Hence, the
presence of initial anisotropy does not adversely affect the
features of the DGP model. It is possible to have an anisotropic
brane in the DGP model and still get a self-accelerating universe
in the late time, as the anisotropic term appearing in
(\ref{late1}) is not like energy density. Also, for a large value
of $S$ ($S\rightarrow\infty$), the anisotropic term tends towards
zero and we recover the isotropic case.

It can be noticed that in the present work, ghost like
instabilities in the self-acclerated branch, as pointed out by
\cite{ant}, do not appear. This may be due to the fact that we are
using special boundary conditions. So, it is interesting to study
the self-accelerated branch with a ghost term, by solving the full
brane-bulk system and considering more realistic boundary
conditions, which is beyond the scope of the present paper.

\end{document}